# Growing small-world networks based on a modified BA model*

XU Xin-Ping(徐新平) [1][**], LIU Feng(刘峰)[1], LI Wei(李炜)[2]

[1]*Institute of Particle Physics, HuaZhong Normal University, Wuhan 430079, China*

[2]*Max Planck Institute for Mathematics in the Sciences, Inselstr.22-26, D-04103, Leipzig, Germany*

We propose a simple growing model for the evolution of small-world networks. It is introduced as a modified BA model in which all the edges connected to the new nodes are made locally to the creator and its nearest neighbors. It is found that this model can produce small-world networks with power-law degree distributions. Properties of our model, including the degree distribution, clustering, and the average path length are compared with that of the BA model. Since most real networks are both scale-free and small-world networks, our model may provide a satisfactory description for empirical characteristics of real networks.

*PACS*: *89.75.-k, 89.75.Fb, 89.75.Hc, 02.50.Cw*

## 1. Introduction

Networks of interacting entities or agents such as humans, computers, animal species, proteins, and neurons have been investigated vigorously.[1] These real networks are equipped with short average path length, high clustering, and scale-free topology.[2-4] The small-world property denotes that the average shortest path length $L$ between vertex pairs in a network grows logarithmically with network size $N$. The clustering structure is characterized by the clustering coefficient $C$ which is the fraction of pairs between the neighbors of a vertex that are directly connected to each other. The high degree of clustering indicates that if vertices $A$ and $B$ are linked to vertex $C$, then $A$ and $B$ are also likely to be linked to each other. It is clarified that networks with high clustering have hierarchical organization and modular structure.[5-7] Networks are called scale-free networks if the degree follows a power-law distribution. In the physics literature, networks with high clustering are commonly modeled by the small-world network model of Watts and Strogatz (WS model), while networks with the power-law degree distribution by the scale-free network model of Barabá si and Albert (BA model).[1] Although both models have a small average path length, each model lacks the property of the other model: the WS model shows a high clustering but without the power-law degree distribution, while the BA model with the scale-free nature does not possess the high clustering. In fact, lots of empirical researches show that most real-world networks both have the small-world and scale-free structure. This has inspired the proposal of more perfect models, aiming at producing these features simultaneously.

The simplest clustered scale-free network model was proposed by Dorogovtsev, Mendes and Samukhin(DMS model).[8] In this model, the new vertices are connected to both ends of a randomly chosen link by two undirected links. The DMS model is insufficient to describe real networks since it has a fixed average degree.[9] In Ref. [10], Holme and Kim extend the BA model to include a "triad formation step" (HK model). The HK model has both the perfect power-law degree distribution and high clustering controlled by the average number of triad formation trials per time step. Therefore, the HK model provides an alternative possibility to achieve the high clustering and the scale-free nature. Recently, in order to study other mechanisms for forming small-world networks, Ozik, Hunt and Ott have introduced a simple evolution model of growing small-world networks (OHO model),[13] in which all connections are made locally to geographically nearby sites. The OHO model shows a high average clustering but has an exponentially decaying degree distribution. More efforts of growing scale-free networks with high clustering are summarized in Ref. [10-12].

In this letter, we introduce a modified BA model that generates small-world networks with scale-free properties. The BA model is based on a simple principle of preferential attachment. The probability that an old node receiving links is linearly proportional to its degree. This assumes that every new node has the complete information about the whole network, which is unrealistic for real network formations. [14, 15] As a matter of fact, many networks have their topology influenced by the environmental constraints. The nodes are separated by some physical distance and thus their ability to know the complete state of all the network nodes at a given time is restricted. In this sense, we assume that the attachments of links of the new nodes are made locally to the creator and its surroundings in our model. That is, when a new node appears, it forms links only to the creator and its nearest neighbors. As we will see, this general mechanism accounts for the emergence of small-world property.

* Supported by NSFC under projects 10375025, 10275027 and by the MOE under project CFKSTIP-704035.

**Email: xuxp@iopp.ccnu.edu.cn

## 2. The model

Let us introduce our model in detail. See Fig.1, Initially, the network is composed of $m_0$ ( $m_0 \geq 2$ )fully connected nodes. At each subsequent time, we grow the network according to the following prescription: (a) At each increment of time, a new node $n$ is built or created by one of the existing vertices, and assume the probability that the new node established by node $i$ is proportional to the degree $k_i$, more precise, the probability of node $i$ to build new node at one step time is

$$p_{cr}(i) = \frac{k_i}{\sum_j k_j}. \quad (1)$$

This equation indicates that nodes with large degrees have large probability to build new nodes. (b) Once the new node is established by node $i$, $m = m_0$ edges are distributed to $i$ and its nearest neighbors.[16] One edge connects node $i$ and the other $(m-1)$ edges are attached to the nearest neighbors. Here we assume the $(m-1)$ links are distributed to the nearest neighbors of node $i$ uniformly, thus the conditional probability that one of the nearest neighbors $j$ connecting the new node established by node $i$ is

$$p(j|i) = \frac{m-1}{k_i} \quad (2)$$

These steps are repeated sequentially, creating a network with a temporally growing number of nodes $t$. We note that, since the network size $t$ is increased by one in each discrete time step, $t$ can be used as a system size or a time variable.

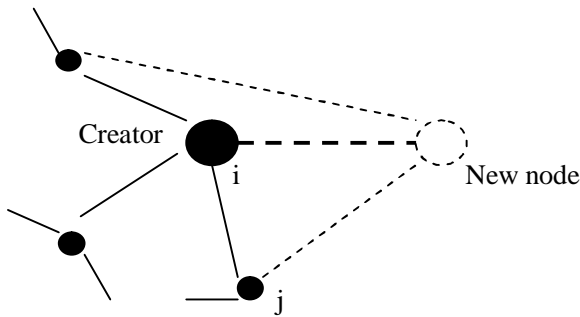

FIG. 1. Illustration of our growing network model. Once a new node is built by the creator $i$, all the edges of the new node are made locally to the creator and its nearest neighbors (dashed lines).

## 3. Degree distribution

We now consider the dynamics of degree $k_i$ of a given node $i$ to calculate the degree distribution in our model. The degree $k_i$ will increase when a new node is established by node $i$ or by its nearest neighbors. Consequently, $k_i$ satisfies the following dynamical equation,

$$\frac{dk_i}{dt} = 1 \bullet p_{cr}(i) + \sum_{l \in \Gamma(i)} p(i|l) \bullet p_{cr}(l) \quad (3)$$

Where $\Gamma(i)$ is the set of nearest neighbors of node $i$. The first term on the right-hand side corresponds to the random selection of node $i$ as a creator of the new node with probability $p_{cr}(i) = \frac{k_i}{\sum_j k_j}$, while the second term corresponds to the selection of the nearest neighbors of node $i$ as a creator. Substituting $p_{cr}(l) = \frac{k_l}{\sum_j k_j}$ and $p(i|l) = \frac{m-1}{k_l}$ into Eq. (3), we have

$$\frac{dk_i}{dt} = m \frac{k_i}{\sum_j k_j} \quad (4)$$

Eq. (4) indicates that the growth and linear preferential attachment are preserved in our model. This equation can be solved with the initial condition $k_i(t = t_i) = m$, yielding

$$k_i(t) = m\sqrt{\frac{t}{t_i}} \quad (5)$$

The evolution of the degree $k_i$ is the same as that in the BA model, therefore the degree distribution of our model is identical to that of the BA model, [17] as well as the HK model. Fig.2 (a) shows $k_i$ of our model versus that of the BA model for $t = 10000$ and $m = 2$, while Fig.2 (b) shows the degree distribution of our model and the BA model. One may think our model is the same as the BA model, but further analytic results and simulation show that our model and the BA model are different in some aspects and our model may be closer to reality than the BA model.

## 4. Clustering and average path length

The clustering $C_i$ of node $i$ is defined by $C_i = E_i/[(1/2)k_i(k_i-1)]$, where $E_i$ is the total number of links between the $k_i$ neighbors of node $i$, and $k_i(k_i-1)/2$ is the maximum number of links that could exist between $k_i$ nodes. We can calculate the clustering of our model exactly for $m=2$, the lower limit of the clustering is given for $m>2$.

When a new node is established at time $t$, the total number of edges $E_i(t)$ between the nearest neighbors of node $i$ can increase either if the new node is built by $i$ or by its nearest neighbors. The evolution equation for $E_i$ is thus given by

$$\frac{dE_i(t)}{dt} = (m-1)\frac{k_i}{\sum_j k_j} + \sum_{l\in\Gamma(i)} p(i|l)\bullet p_{cr}(k_l)$$

$$= 2(m-1)\frac{k_i}{\sum_j k_j} \quad (6)$$

Considering that there are at least $m-1$ links among the nearest neighbors of node $i$ when it is established, the initial condition of $E_i$ can be written as $E_i(t=i) \geq (m-1)$, where the equality holds only for $m=2$. Therefore Eq. (6) can be readily integrated as

$$E_i(t) \geq \frac{2(m-1)k_i(t)}{m} - (m-1) \quad (7)$$

For the particular case of $m=2$, $E_i(t) = k_i(t)-1$, the clustering coefficient of node $i$ is given by

$$C_i(t) = \frac{2E_i(t)}{k_i(t)(k_i(t)-1)} = \frac{2}{k_i(t)} \quad (8)$$

Eq. (8) indicates that the clustering spectrum scales as $C_i(k_i) \sim k_i^{-1}$, which is similar to that observed in Ref.[18,19]. Fig.3. shows the clustering spectrum for $t=10000$, which agrees well with the analytical prediction of (8). Noting that the degree distribution is $p(k) = \frac{12}{k(k+1)(k+2)}$,[1] the clustering coefficient C of the whole network can be given by

$$C = 2\sum_{k=2}^{\infty}\frac{p(k)}{k} = 2\sum_{k=2}^{\infty}\frac{12}{k^2(k+1)(k+2)} \approx 0.7392 \quad (9)$$

This is a large value, comparable, for instance, with the clustering coefficient for the collaboration network of movie actors C=0.79. [20] Through our numerical simulation, we obtain a value of C=0.74 for $t=10000$, which is close to the prediction of Eq.(9). For larger $t$, this value is observed to approach the theoretical result. In networks with larger values of $m$ we also observe approach of $C$ to a constant asymptotic value as $t$ increases (e.g., for $m=3$, $C=0.618$). On the contrary, as $t$ grows, the clustering coefficient of the BA model scales as $\ln^2 t/t$,[21] implying that the clustering vanishes as $t\to\infty$. We note that many real networks have a high clustering, and thus believe that our model is more realistic.

The open circles in Fig. 4 show L, the shortest path length between pairs of nodes averaged over all node pairs of single growing network realizations, on a linear scale versus $t$ on a logarithmic scale. The data shows a linear trend, demonstrating the average path length increase logarithmically with the network size $t$. In addition, from Fig.4, we can conclude that the local restriction of edges make the average path length of our model larger than that of the BA model.

Since the network has a high clustering as $t\to\infty$ and the average path length scales as $L\sim\ln t$, our model displays the small-world properties.

## 5. Conclusion and discussions

In conclusion, we present a small-world network model with power-law degree distribution. This model provides a physically realistic mechanism in which the establishment of the new node is affected by a certain existing node and its nearest neighbors, and therefore the appearance of each individual node influenced by a certain entity and its surroundings, can form networks with small-world characteristics.

We would like to point out that, while the HK model based on the triad formation is probably a reasonable model for how some small-networks are formed, our model provides an alternative possible mechanisms for forming small-world networks.[22] The novel growth of old node setting up or giving birth to a new node incarnates the global preferential attachment, while the restriction that all the edges locally distributed to the nearest neighbors implies the local attachment preference. This is the familiar case that the new entity not only gets interaction with the old entity who creates it, but also get interactions with the partners or fellows of the creator. Take the scientists collaboration networks for example,[23,24] a scientist joining a research group will often collaborate with both the group director (the creator) and the other group members (neighbors). The scientist's affiliation can naturally boost the research productivity of the group and also, enhance the collaborations of other members. [25] Therefore, our model based on the global and local preferential attachment, suggest that small-world networks in environmental constrained physical systems may be a natural consequence of system growth and environmental interactions.

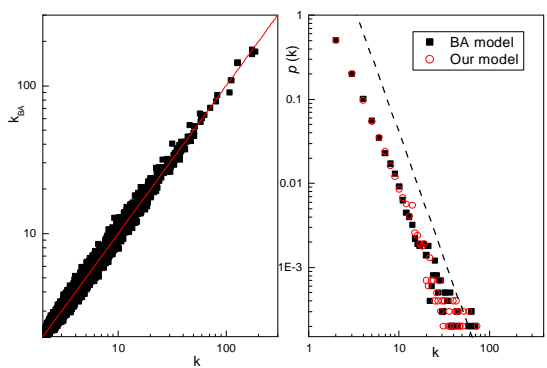

Fig.2. (a) Degree $k_i$ of our model versus that of the BA model for $t = 10000$ and $m = 2$, the simulation results are averaged over 100 realizations, the solid line is the bisector of the first quadrant. (b) Degree distributions for our model and BA model, the dashed line is a power-law $p(k) \sim k^{-3}$.

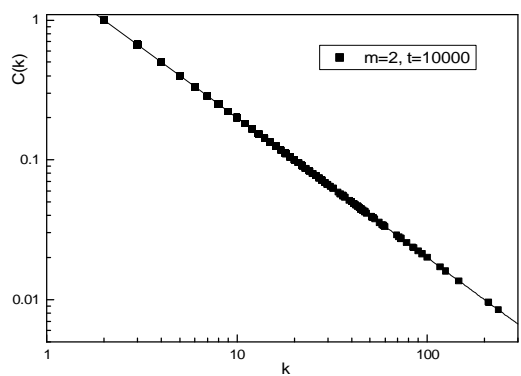

Fig.3. The degree dependent local clustering coefficient for $m = 2$ and $t = 10000$, the solid line is the analytical prediction of Eq.(8). The simulation results agrees well with the analytical prediction $C(k) = \frac{2}{k}$

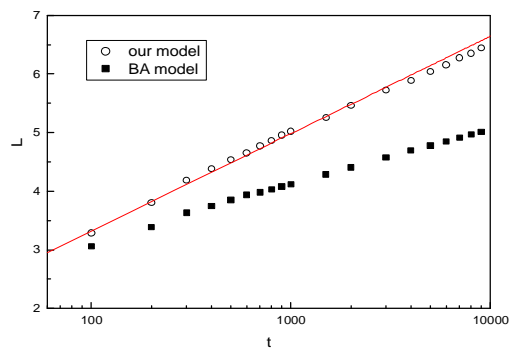

Fig.4. Semi logarithmic graph of the average path length $L$ vs the system size $t$ for our model (circles) and BA (squares) model. The data indicates that our model shows the small-world slow path length growth characteristic, $L \sim \ln t$. The straight line is the function $L = \ln t / \ln <k> = \ln t / \ln 4$.